\documentstyle[12pt,preprint,aps,epsfig]{revtex}
\tighten
\newcommand{\beq}{\begin{eqnarray}}
\newcommand{\eeq}{\end{eqnarray}}
\newcommand{\non}{\nonumber\\ }

\newcommand{\bsga}{ B\to X_s \gamma }
\newcommand{\bsgg}{ b\to s \gamma \gamma }

\newcommand{\bstogg}{B_s \to \gamma \gamma }

\newcommand{\bsdtogg}{B_{s,d} \to \gamma \gamma }

\newcommand{\calb}{ {\cal B} }
\newcommand{\lqcd}{ \Lambda_{QCD} }
\newcommand{\rcppm}{ r^\pm_{CP} }

\newcommand{\rcpm}{ r^-_{CP} }

\newcommand{\tab}[1]{Table \ref{#1}}

\newcommand{\fig}[1]{Fig. \ref{#1}}
\newcommand{\figs}[1]{Figs. \ref{#1}}

\newcommand{\paa}{\pi_1^{\pm}}
\newcommand{\pbb}{\pi_8^{\pm}}
\newcommand{\pcc}{\tilde{\pi}^{\pm}}
\newcommand{\pcm}{\tilde{\pi}^-}

\newcommand{\pcz}{\tilde{\pi}^0}

\newcommand{\fq}{F_Q  }

\newcommand{\mw}{ m_W }
\newcommand{\mpaa}{m_{\pi_1}}
\newcommand{\mpbb}{m_{\pi_8}}
\newcommand{\mpcc}{m_{\tilde{\pi}}}
\newcommand{\mpdd}{m_{\tilde{H}}}

\def \cpl{ {Chin. Phys. Lett.}  }
\def \ctp{ {Commun. Theor. Phys.}  }
\def \epjc{{Eur. Phys. J. C} }

\def \mpla{{Mod. Phys. Lett. A} }
\def \npb{ {Nucl. Phys. B} }
\def \plb{ {Phys. Lett. B} }
\def \prd{ {Phys. Rev. D} }
\def \prl{ {Phys. Rev. Lett.}  }

\def \rmp{ {Rev. Mod. Phys. }  }
\def \prl{ {Phys. Rev. Lett.}  }
\def \prl{ {Phys. Rev. Lett.}  }
\def \zpc{ {Z. Phys. C }  }

\title{ Technicolor corrections on $B_{s,d} \to \gamma\gamma$ decays
in QCD factorization }
\author{ Zhenjun Xiao \\
{\small Physics Department, Nanjing Normal University,
Nanjing, Jiangsu 210097, China}\\
{\small and CCAST(World Laboratory), P.O.Box 8730, Beijing
100080, China }\\
\vspace{0.5cm}
Cai-Dian  L\"{u} \\
{\small CCAST(World Laboratory), P.O.Box 8730, Beijing
100080, China }\\
{\small and  Institute of High Energy Physics, CAS, P.O. Box
918(4), Beijing 100039, China}\\
\vspace{0.5cm}
Wujun Huo \\
{\small Ottawa-Carleton Institute for Physics, Department of
Physics, Carleton University, Ottawa, Canada}\\
{\small and Department of Physics, Peking University, Beijing
100871, China } }
\date{\today}
\begin{document}
\maketitle
\begin{picture}(0,0)(0,0)
\put(310,250){{\large BIHEP-TH-2002-14}}
\end{picture}
\begin{abstract}
Within the framework of the Top-color-assisted Technicolor (TC2) model, we calculate the
new physics contributions to the branching ratios $\calb(B_{s,d} \to \gamma \gamma)$
and CP violating asymmetries $\rcpm(B_{s,d} \to \gamma \gamma )$
in the QCD factorization based on the heavy-quark limit $m_b \gg \Lambda_{QCD}$.
Using  the considered parameter space, we find that
(a) for both $B_s\to \gamma \gamma$ and $B_d \to \gamma \gamma$ decays, the new
physics contribution can provide a factor of two to six enhancement to their
branching ratios,
(b) for the $B_s \to \gamma \gamma$ decay, its direct CP violation is very small
in both the SM and TC2 model, and
(c) the CP violating asymmetry $\rcpm(B_d \to \gamma \gamma)$ is around the
ten percent level in both the SM and TC2 model, but the sign of CP asymmetry
in the TC2 model is different from  that in the SM.
\end{abstract}


\pacs{PACS numbers: 13.25.Hw, 12.60.Nz, 14.40.Nd}

\newpage
\section{Introduction}

As is well known, the rare radiative decays of B mesons induced by the quark
decay  $b \to q \gamma$ ($q=d,s$) are very sensitive to the flavor structure
of the standard model (SM) and to new physics beyond the SM.
Both inclusive and exclusive processes, such as the decays $B \to X_s \gamma$,
$X_s \gamma \gamma$ and $B_{s,d} \to \gamma \gamma$, have been studied in great detail
\cite{yao,cmm97,buras98,kagan99,neubert02,hiller,hiller98,bosch02a,bosch02b,xiao95,xiao00}.

The inclusive decay $ B  \to X_s \gamma $ is measured experimentally
with increasing accuracy\cite{bsgaexp}. The world average as given by the 2002 Particle
Data Group ( PDG2002 ) \cite{pdg2002} is
\beq
\calb (\bsga) = (3.3 \pm 0.40) \times 10^{-4}, \label{eq:bsgaexp}
\eeq
which is quite consistent with the next-to-leading order (NLO) standard model prediction
\cite{kagan99}
\beq
\calb (\bsga)^{TH} = (3.29 \pm 0.34) \times 10^{-4}, \label{eq:bsgath}
\eeq
Obviously, there is only  small room left for new physics effects in
flavor-changing neutral current processes based on  the $b \to s$ transition.
In other words, the excellent agreement between SM theory and experimental data results
in a strong constraint on many new physics models beyond the SM.

Within the SM, the electroweak contributions to $\bsgg $ and $B \to \gamma\gamma$
decays have been calculated some time ago\cite{yao}; the
leading-order QCD corrections and the long-distance contributions were evaluated
recently by several groups\cite{hiller,long}. The new physics corrections
were also considered, for example, in the two-Higgs doublet model \cite{2hd,xiao01} and
the supersymmetric model \cite{susy}.

On the experimental side, only  upper limits ($90\%C.L.$) on the branching ratios of
$B_{s,d} \to \gamma \gamma$  are currently available
\beq
\calb (B_s\to \gamma \gamma ) &<& 1.48 \times 10^{-4}, \cite{cleo95}\label{eq:cleo95} \\
\calb (B_d\to \gamma \gamma ) &<& 1.7 \times 10^{-6}, \cite{babar01}\label{eq:babar01}
\eeq
which are roughly two orders above the SM predictions \cite{yao,hiller,bosch02a,bosch02b}.
These radiative decays are indeed very interesting because
(a) these decays have a very clean signal where two monochromatic energetic photons
are produced precisely back-to-back in the rest frame of B meson;
(b) these exclusive decays also allow us to study the CP violating
effects as the two photon system can be in a CP-even or CP-odd state;
(c) since $\bstogg $ depends on the same set of Wilson coefficients as $B \to X_s \gamma$,
its sensitivity to new physics beyond the SM complements the corresponding
sensitivity in $B \to X_s \gamma$; and
(d) the smallness of the branching ratios can be compensated
by the very high statistics expected at the current B factory experiments
and future hadron colliders.

In this paper, we present our calculation of branching ratios
and CP-violating asymmetries for rare exclusive decays $\bsdtogg$ in the framework of
the Top-color-assisted Technicolor (TC2) model \cite{hill95} by employing
the QCD factorization based on the heavy-quark limit $m_b \gg \Lambda_{QCD}$
\cite{bosch02a,bosch02b}.

This paper is organized as follows: In Sec. II, we give a brief review
about the SM predictions for the branching ratios and CP asymmetries of
$B_{s,d} \to \gamma \gamma $ decays. In Sec.~III, we present the
basic ingredients of the TC2 model, and evaluate the new penguin diagrams.
After studying  the constraint on the TC2 model by considering the data of
$B_d^0-\bar{B}^0_d$ mixing and $\bsga$ decay, we find the Wilson coefficients
$C_7$ and $C_8$ with the inclusion of the new physics (NP) contribution.
In Sec.~4,  we show the numerical results
of branching  ratios  and CP-violating asymmetries for $\bsdtogg$ decays.
The discussions and conclusions are included in the final section.

\section{$B_{s,d} \to \gamma \gamma$ decays in the SM} \label{sec:sm}

In this section, based on currently available studies,  we present the formulae
for exclusive decay $\bsdtogg$ in the framework of SM.

\subsection{Effective Hamiltonian for inclusive $b\to s \gamma \gamma $ decay}

We know that the quark level processes $b \to s \gamma,  s \gamma \gamma$ and the exclusive
decays $\bsdtogg$ have a close relation. Up to the order $1/m^2_W$, the effective
Hamiltonian for the decay $b\to s \gamma \gamma$ is identical to the one for $\bsga$
transition \cite{yao,hiller98}
\beq
{\cal H}_{eff} (b \to s \gamma) = {\cal H}_{eff} (b \to s \gamma \gamma) + {\cal
O}\left (\frac{1}{\mw^2}\right ) \label{eq:heffa}.
\eeq
This can be understood by either applying the equation of motion \cite{simma94}
or by applying an extension of Low's low energy theorem\cite{low}.

Up to corrections of order $1/m_W^2$, the effective Hamiltonian for
$b \to s\gamma \gamma$ is just the one for $b \to s \gamma$ and takes the form
\beq
{\cal H}_{eff}=\frac{G_F}{\sqrt{2}}\sum_{p=u,c}\lambda_p^{(s)}
\left[ C_1(\mu) Q^p_1 + C_2(\mu) Q^p_2 +\sum_{i=3,\ldots ,8} C_i(\mu)
Q_i\right], \label{eq:heff}
\eeq
where $\lambda_p^{(s)}=V^*_{ps}V_{pb}$ is the Cabibbo-Kobayashi-Maskawa (CKM)
factor. And the current-current,
QCD penguin,  electromagnetic and chromomagnetic dipole operators are  given by
\footnote{For the numbering of operators $Q_{1,2}^p$, we use the convention of
Buras {\it et al.} \cite{buras98} throughout this paper.}
\beq
 Q^p_1 &=& (\bar s_\alpha p_\beta)_{V-A}(\bar p_\beta b_\alpha)_{V-A}, \\
 Q^p_2 &=& (\bar s p)_{V-A}(\bar p  b)_{V-A},  \\
 Q_3 &=& (\bar s b)_{V-A} \sum_{q=u,d,s,c,b} (\bar q q)_{V-A}, \\
 Q_4 &=& (\bar s_\alpha b_\beta)_{V-A} \sum_{q=u,d,s,c,b} (\bar q_\beta q_\alpha)_{V-A}, \\
 Q_5 &=& (\bar sb)_{V-A} \sum_{q=u,d,s,c,b} (\bar q q)_{V+A}, \\
 Q_6 &=& (\bar s_\alpha b_\beta)_{V-A} \sum_{q=u,d,s,c,b} (\bar q_\beta q_\alpha)_{V+A}, \\
 Q_7 &=& \frac{e}{4\pi^2}\,\bar s_\alpha \sigma^{\mu\nu}\left (m_b R + m_s L \right )
                b_\alpha \, F_{\mu\nu}\, ,    \\
 Q_8 &=& \frac{g_s}{4\pi^2}\,\bar s_\alpha \sigma^{\mu\nu}\left (m_b R + m_s L \right )
    T^a_{\alpha \beta} b_\beta\,  G^a_{\mu\nu} \, ,  \label{eq:qi}
\eeq
where $\alpha$ and $\beta$ are color indices, $\alpha=1, \ldots,8$ labels $SU(3)_c$
generators,  $e$ and $g_s$ refer to the electromagnetic and strong coupling constants,
and $L,R =(1\mp \gamma_5)/2$, while $F_{\mu\nu}$ and $G^a_{\mu\nu}$
denote the photonic and gluonic field strength tensors, respectively. In $Q_{7,8}$, the terms
proportional to $m_s$ are usually neglected because of the strong suppression $m_s^2/m_b^2$.
The effective Hamiltonian for $b \to d \gamma \gamma$ is obtained from
Eqs.(\ref{eq:heff}) - (\ref{eq:qi}) by the replacement $s \to d$.

The Wilson coefficients $C_i(\mu)$ in Eq.(\ref{eq:heff}) are known currently at next-to-leading
order (NLO) \cite{cmm97,buras98}. Within the SM and at scale $\mw$,
the Wilson coefficients $C_i(\mw)$ at the leading order (LO)
approximation have been given for example in \cite{buras98},
\beq
C_i(\mw ) &= & 0 \ \ \ (i=1,3,4,5,6), \label{eq:c1mw}\\
C_2(\mw)  &= & 1, \label{eq:c2mw}, \\
C_7(\mw) &=& \frac{-7 x_t + 5x_t^2 + 8x_t^3}{ 24(1-x_t)^3}
    - \frac{ 2 x_t^2-3 x_t^3 }{ 4(1-x_t)^4}\log[x_t], \label{eq:c7mw}\\
C_8(\mw) &=& \frac{-2 x_t - 5x_t^2 + x_t^3}{ 8(1-x_t)^3}
- \frac{ 3 x_t^2 }{ 4(1-x_t)^4}\log[x_t], \label{eq:c8mw}
\eeq
where $x_t=m_t^2/\mw^2$.

By using QCD renormalization group equations\cite{buras98}, it is
straightforward to run Wilson coefficients $C_i(\mw)$ from the scale $\mu ={\cal O}( \mw)$
down to the lower scale $\mu ={\cal O}(m_b)$. The leading order results for the Wilson coefficients
$C_i(\mu)$ with $\mu \approx m_b$ are of the form \cite{buras98}
\beq
C_j(\mu) &=& \sum_{i=1}^8 k_{ji} \eta^{a_i}\ \ \ (j=1,\ldots, 6), \label{eq:c1mu}\\
C_7(\mu) &=& \eta^{16/23} C_7(\mw) +\frac{8}{3} \left (  \eta^{14/23} - \eta^{16/23} \right )
     C_8(\mw) + \sum_{i=1}^8 h_i \eta^{a_i}\, , \label{eq:c7mu}\\
C_8(\mu) &=& \eta^{14/23} C_8(\mw) + \sum_{i=1}^8 \bar{h}_i \eta^{a_i}\, , \label{eq:c8mu}
\eeq
where  $\eta=\alpha_s(\mw)/\alpha_s(\mu)$, and the magic numbers are \cite{buras98}
\beq
a_i&=& ( 14/23,16/23,6/23,-12/23,0.4086, -0.4230, -0.8994,0.1456)\, , \label{eq:ai}\\
h_i &=& ( 2.2996,-1.0880,-3/7,-1/14,-0.6494,-0.0380,-0.0185,-0.0057)\, ,  \label{eq:hi}\\
\bar{h}_i &=& (0.8623, 0,        0,    0,-0.9135, 0.0873,-0.0571, 0.0209)\, , \label{eq:hibar}\\
k_{ji}& =&  \left( \begin{array}{cccc cccc}
      0&0&1/2  &-1/2& 0     & 0     & 0     & 0     \\
      0&0&1/2  & 1/2& 0     & 0     & 0     & 0     \\
      0&0&-1/14& 1/6& 0.0510&-0.1403&-0.0113& 0.0054\\
      0&0&-1/14&-1/6& 0.0984& 0.1214& 0.0156& 0.0026\\
      0&0&0    & 0  &-0.0397& 0.0117&-0.0025& 0.0304\\
      0&0&0    & 0  & 0.0335& 0.0239&-0.0462&-0.0112\\
\end{array} \right) \quad .
\eeq

The numerical results of the LO Wilson coefficients  $C_i(\mu)$ obtained by using  the input
parameters as given in \tab{input} are listed in \tab{cieff} for $\mu =m_b/2, m_b$ and $2 m_b$,
respectively.

\subsection{$B_{s,d} \to \gamma \gamma$ decays in the SM}

Based on the effective Hamiltonian for the quark level process $b \to s(d)
\gamma \gamma$, one can write down the amplitude for $B_{s,d} \to \gamma \gamma$ and calculate
the branching ratios and CP violating asymmetries once a method is derived for computing
the hadronic matrix elements.
There exist so far two major approaches for the theoretical treatments of
exclusive decay $B \to \gamma \gamma$.

The first approach was proposed ten years ago and has been employed by many authors
\cite{yao,hiller}. Under this approach, one simply evaluates the hadronic
element of the amplitudes for one-particle reducible (1PR) and one-particle
irreducible (1PI) diagrams, relying on a phenomenological model.
One can work, for example, in the weak binding approximation and assume that both the $b$
and the light $q$ quarks are at rest in the $B_q$ meson\cite{reina97}. From the heavy quark
effective theory(HQET), for instance, one can also assume that the velocity of the
$b$ quark coincides with the velocity of the $B_q$ meson up to a residual momentum
of $\Lambda_{QCD}$. Both pictures are compatible up to corrections of order ($\Lambda_{QCD}/m_b$)
\cite{reina97}. One typical numerical result obtained by employing this approach is
\beq
\calb (B_s \to \gamma \gamma) \approx (2 -8 )\times 10^{-7}
\eeq
after inclusion of LO QCD corrections \cite{reina97}.
There are also many works concerning the estimation
of the long distance contributions to $B \to \gamma \gamma$ decay\cite{long}.

In the first approach, one has to employ hadronic models to describe the
$B_q$ ($q=s,d$)  meson bound state dynamics. It is thus impossible for one to
separate clearly the short- and long-distance dynamics and to make distinctions
between the model-dependent and model-independent features.

The second approach was proposed recently by Bosch and Buchalla\cite{bosch02a,bosch02b}.
They analyzed the $B_{s,d} \to \gamma \gamma $ decays in QCD factorization
approach based on the heavy quark limit $m_b \gg \lqcd$. Under this approach,
one can systematically separate perturbatively calculable hard scattering
kernels from the nonperturbative B-meson wave function. Power counting in $\lqcd/m_b$
allows one to identify leading and subleading contributions to $B \to \gamma
\gamma$. In this paper, we will employ the Bosch and Buchalla (BB) approach to calculate
the Technicolor corrections to $B_{s,d} \to \gamma \gamma$ decays.

\begin{figure}[t]    
\vspace{-150pt}
\begin{minipage}[]{\textwidth}
\centerline{\epsfxsize=\textwidth \epsffile{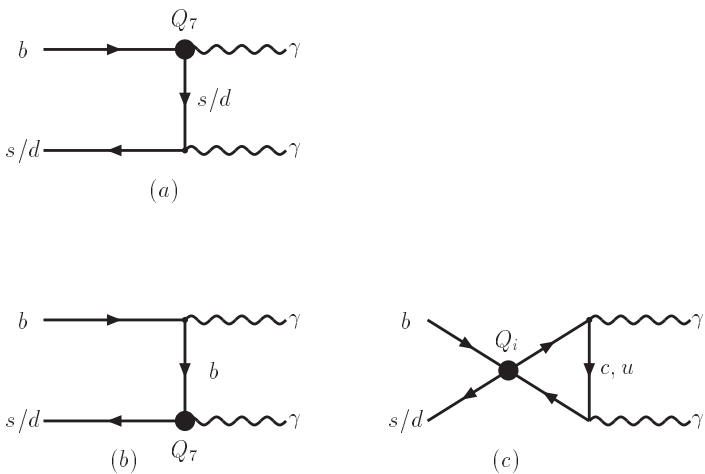}}
\vspace{-250pt}
\caption{ The leading power 1PR diagram (a) and subleading power 1PR diagram (b) of
the magnetic penguin operator $Q_7$, and the subleading power 1PI diagram (c) of the
four-quark operators $Q_i$. The diagrams with interchanged photons are not shown. }
\label{fig:fig1}
\end{minipage}
\end{figure}

From Refs.\cite{bosch02a,bosch02b}, one knows that
(a) only one 1PR diagram [ \fig{fig:fig1}(a) ]  contributes at leading power;
(b) the most important subleading contributions induced by the 1PR [\fig{fig:fig1}(b) ]
and 1PI diagrams [\fig{fig:fig1}(c) ] can also be calculated; and (c) the direct CP violation of
$B_d \to \gamma \gamma $   can reach the $10\%$ level.

The amplitude for the $B \to \gamma \gamma$ decay has the general
structure\cite{bosch02a}
\beq
{\cal A}(\bar B\to\gamma(k_1,\epsilon_1)\gamma(k_2,\epsilon_2)) =
 \frac{G_F}{\sqrt{2}} \frac{\alpha_{em}}{3\pi} f_B
 \frac{1}{2}\langle\gamma\gamma |  A_+ F_{\mu\nu} F^{\mu\nu}
 -i\, A_- F_{\mu\nu} \tilde F^{\mu\nu}|0 \rangle .
 \label{eq:a-bgg}
\eeq
Here $F^{\mu\nu}$ and $\tilde F^{\mu\nu}=\varepsilon^{\mu\nu\lambda\rho}F_{\lambda\rho}/2 $
are the photon field strength tensor and its dual with $\varepsilon^{0123}=-1$.
The branching ratio of $B_q \to \gamma \gamma $ decay  with $q= s,d$ is then given by
\beq
{\calb }(\bar{B}_q \to \gamma \gamma ) =
\tau_{B_q}\, \frac{G^2_F m^3_{B_q} f^2_{B_q} \alpha_{em}^2}{288\pi^3}\left(|A_+|^2 +
|A_-|^2\right)\label{eq:gamma-bgg}
\eeq
where $G_F$ is the Fermi constant, $\alpha_{em}$ is the fine structure constant,
$\tau_{B_q}$ is the lifetime of $B_q$ meson, and $m_{B_q}$ and $f_{B_q}$ are the mass and
decay constant of the $B_{q}$ meson, respectively. The values of all input parameters are
listed in \tab{input}.

The matrix elements of the operators $Q_i$ in Eq.(\ref{eq:heff}) can be written
as
\beq
\langle \gamma(\epsilon_1)\gamma(\epsilon_2)|Q_i|\bar B\rangle = f_B
\int^1_0 d\xi\, T^{\mu\nu}_i(\xi)\, \Phi_B(\xi)\epsilon_{1\mu} \epsilon_{2\nu}
\eeq
where the $\epsilon_i$ are the polarization 4-vectors of the photons,
$\Phi_B\equiv\Phi_{B1}$ is the leading twist light-cone distribution
amplitude of the $B$ meson, and $ T^{\mu\nu}_i(\xi)$ is the hard-scattering kernel
describing the hard-spectator contribution.

By explicit calculations as were done in Ref.\cite{bosch02a}, the quantities $A_{\pm}$
in Eq.(\ref{eq:gamma-bgg}) are of the form
\beq
A_+ &=& \lambda_{u}^{(q)} A_+^u +  \lambda_{c}^{(q)} A_+^c, \label{eq:aplus}\\
A_- &=& \lambda_{u}^{(q)} A_-^u +  \lambda_{c}^{(q)} A_-^c, \label{eq:aminus}
\eeq
with
\beq
A^p_+ &=& -C_7 \frac{m_{B}}{\lambda_{B}} + \left (C_5 + 3 C_6 \right )
\left [ \frac{1}{2} g(1) -\frac{1}{3}\right ],   \label{eq:app} \\
A^p_- &=& -C_7 \frac{m_{B}}{\lambda_{B}} - \frac{2}{3}\left ( C_2+ 3 C_1 \right )g(z_p)
-\left (C_3 -  C_5 \right ) \left [ 2 g(z_c) + \frac{5}{6}g(1)\right ]\non
&& -\left (C_4 -  C_6 \right ) \left [ \frac{2}{3} g(z_c) + \frac{7}{6}g(1)\right ]
+\frac{20}{3}C_3 +4 C_4 - \frac{16}{3}C_5 ,  \label{eq:apm}
\eeq
where $z_p= m_p^2/m_b^2$ for $p = u,c$, and
\beq
 g(z)=-2+4z\left[Li_2\left(\frac{2}{1-\sqrt{1-4z+i\epsilon}}\right)+
       Li_2\left(\frac{2}{1+\sqrt{1-4z+i\epsilon}}\right)\right]\label{eq:gz}
\eeq
 and $ Li_2(x)$ is the dilogarithm function. It is easy to see that $A^u_+ =
 A^c_+$, but $A^u_- \neq  A^c_-$. The function $g(z)$ has an imaginary part for
$0 < z < 1/4$, while  $g(0)=-2$ and $g(1)=2(\pi^2-9)/9$.

The first term of $A_\pm^p$ is the
leading power contribution from the 1PR diagram [ Fig.1(a) ] of the penguin operator $Q_7$,
the remaining terms of $A_\pm^p$ represent the subleading contributions
from the 1PR diagram [ Fig.1(b) ] with the operator $Q_7$ where the second photon
is emitted from the $b$ quark line, and from the 1PI
diagram [ Fig.1(c) ] induced by insertion of four-quark operators
$Q_i$. From the formulas as given in Eq.(\ref{eq:gamma-bgg})
and Eqs.(\ref{eq:aplus}) - (\ref{eq:apm}), we find the numerical results of the
branching ratios  in SM
\beq
{\calb }(\bar{B}_s \to \gamma \gamma ) &=& \left [1.2 ^{+2.4}_{-0.6}(\Delta \lambda_B)
^{+0.3}_{-0.2} (\Delta \mu) \pm 0.3 (\Delta f_{B_s} )
 \pm 0.02(\Delta \gamma) \right ] \times 10^{-6},
\label{eq:brbssm}\\
{\calb }(\bar{B}_d \to \gamma \gamma ) &=& \left [3.2 ^{+6.6}_{-1.6} (\Delta \lambda_B)
^{+0.8}_{-0.6} (\Delta \mu) ^{+1.0}_{-0.9}(\Delta f_{B_d} )
^{+1.1}_{-0.8} (\Delta \gamma)\right ] \times 10^{-8},
\label{eq:brbdsm}
\eeq
where the central values of branching ratios are obtained by using  the central values of
input parameters as given in \tab{input}, and the errors correspond to
$\Delta \lambda_B=\pm 0.15$ GeV, $m_b/2\leq  \mu \leq 2m_b$,
$\Delta f_{B_d}= \Delta f_{B_s}=\pm 0.03$ GeV, respectively. For the CKM angle
$\gamma$, we consider the range of $\gamma = (60 \pm 20)^\circ$ according to the
global fit result \cite{pdg2002}. Obviously, the dominant errors are induced by the uncertainty of
hadronic parameter $\lambda_B$, the renormalization scale $\mu$ and decay constant $f_{B_q}$.
The error induced by $\Delta \gamma$ is about $30\%$ for $B_d$ decay, but very small
for $B_s$ decay. The errors due to the uncertainty of other input parameters are
indeed very small and can be neglected.

Now we consider the CP violating asymmetries of $B_{s,d} \to \gamma \gamma $
decays. Following the definitions of Ref.\cite{bosch02a}, the subscripts $\pm$ on $A_\pm$
for $\bar B\to\gamma\gamma$ decay denote the CP properties of the corresponding two-photon
final states, while   $\bar{A}_\pm$ refer to the CP conjugated amplitudes
for the decay $B\to\gamma\gamma$ (decaying $\bar b$ antiquark).
Then the deviation of the ratios
\beq
r^\pm_{CP}=\frac{|A_\pm|^2-|\bar A_\pm|^2}{|A_\pm|^2+|\bar A_\pm|^2}\label{eq:rcp}
\eeq
from zero is a measure of direct CP violation. Since $A_+^p = \bar{A}_+^p$, $r^+_{CP}$ is
always zero. For $r^-_{CP}$ of $B_d \to \gamma \gamma$ decay, however, it can be rather large.
By using the central values of
input parameters as given in \tab{input} and assuming $\gamma=60^\circ$, we
find
\beq
r^-_{CP} (B_s \to \gamma \gamma) &=& \left [ 0.39 ^{+0.25}_{-0.28}(\Delta \mu)
^{+0.16}_{-0.11}(\Delta \lambda_B) ^{+0.06}_{-0.11}(\Delta \gamma) \right ] \%,
\label{eq:rcpbs}\\
r^-_{CP} (B_d \to \gamma \gamma) &=& \left [ -10.2 ^{+7.3}_{-6.6}(\Delta \mu)
^{+4.3}_{-4.0}(\Delta \lambda_B) ^{+1.4}_{-0.1}  (\Delta \gamma )\right ] \%
\label{eq:rcpbd}
\eeq

It is easy to see that the direct CP violating asymmetry for $B_s \to \gamma
\gamma$ decay is small, $\sim 1\%$, and cannot be detected by experiments.
For $B_d \to \gamma \gamma$ decay. however, its CP violation can be rather
large, around $-10\%$ for $\gamma \sim 60^\circ$. But the much smaller branching
ratio is a great challenge for the current and future experiments.

\begin{figure}[t]       
\vspace{-40pt}
\begin{minipage}[]{\textwidth}
\centerline{\epsfxsize=0.95\textwidth \epsffile{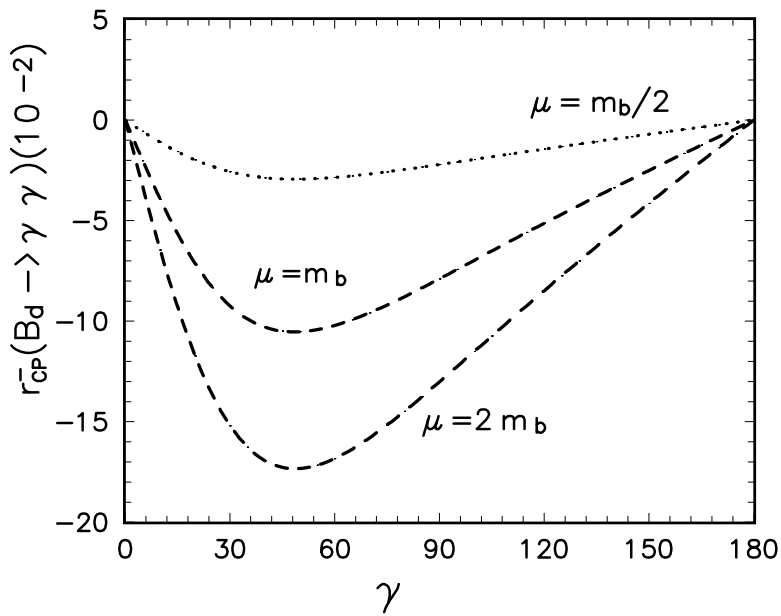}}
\vspace{-30pt}
\caption{The CP violating asymmetry of $(B_d \to \gamma \gamma)$ decay vs the
CKM angle $\gamma$ and energy scale $\mu$ in the SM. The dotted, short-dashed and  solid curves
show the SM predictions for $\mu=m_b/2, m_b$ and $2m_b$, respectively.}
\label{fig:fig2}
\end{minipage}
\end{figure}

In \fig{fig:fig2}, we show the CKM angle $\gamma-$ and $\mu$-dependence of
$r^-_{CP}(B_d \to \gamma \gamma)$. The dotted, short-dashed and solid curves
show the SM predictions of $r^-_{CP}(B_d \to \gamma \gamma)$ for $\mu=m_b/2,
m_b$ and $2m_b$, respectively. The CP violating asymmetry even can reach $-17\%$
for CKM angle $\gamma \approx 50^\circ$, the value preferred by the global fit \cite{global-fit}
and by the analysis based on the measurements of branching ratios of $B \to K\pi$
decays \cite{kpi}. The value of $\rcpm(B_d \to \gamma \gamma)$
here is the same as that given in Ref.\cite{bosch02a} for $\mu=m_b$, but opposite with
what was given in Ref.\cite{bosch02a} for $\mu=m_b/2$ and $2 m_b$,
respectively.

\section{$B_{s,d}\to \gamma\gamma$ decays in TC2 model} \label{sec:tc2model}

In this section, we calculate the loop corrections to $B_{s,d} \to \gamma \gamma$
decays   in TC2 model.

\subsection{TC2 model}

Apart from some differences in group structure and/or particle contents,
all TC2 models \cite{hill95,lane96} have the following common
features:
(a) strong Top-color interactions, broken near 1 TeV, induce a large
top condensate and all but a few GeV of the top quark mass, but contribute
little to electroweak symmetry breaking;
(b) technicolor \cite{weinberg76}  interactions are
responsible for electroweak symmetry breaking, and extended technicolor
(ETC) \cite{etc79} interactions generate
the masses of all quarks and leptons, except that of the top quarks;
(c) there exist top pions $\pcc$ and $\pcz$ with a decay constant
$\fq \approx (40- 50)$ GeV.  In this paper we will chose the well-motivated
and most frequently studied TC2 model proposed by Hill \cite{hill95}
to calculate the contributions to the rare exclusive B decays
from the relatively light charged pseudo-scalars.
It is straightforward to extend the studies in this paper to
other TC2 models.

In the TC2 model\cite{hill95}, after integrating out the heavy coloron and
$Z'$, the effective four-fermion interactions have the
form \cite{buchalla96b}
\beq
{\cal L}_{eff} =\frac{4 \pi}{M_V^2} \left \{
 \left ( \kappa + \frac{2 \kappa_1}{27} \right )
 \overline{\psi}_L t_R \overline{t}_R \psi_L
+  \left ( \kappa - \frac{ \kappa_1}{27} \right )
\overline{\psi}_L b_R \overline{b}_R \psi_L \right \}, \label{eff1}
\eeq
where  $\kappa= (g_3^2/4\pi)\cot ^2\theta$ and
$\kappa_1= (g_1^2/4\pi)\cot ^2\theta'$, and $M_V$ is the mass of
coloron $V^\alpha$ and $Z'$. The effective interactions of Eq. (\ref{eff1}) can
be written in terms of two
auxiliary scalar doublets $\phi_1$ and $\phi_2$. Their couplings to quarks
are given by \cite{kominis95}
\beq
{\cal L}_{eff} = \lambda_1 \overline{\psi}_L \phi_1 \overline{t}_R
+ \lambda_2 \overline{\psi}_L \phi_2 \overline{b}_R, \label{eff2}
\eeq
where $\lambda_1^2 = 4\pi (\kappa + 2\kappa_1/27)$ and
$\lambda_2^2 = 4\pi (\kappa - \kappa_1/27)$. At energies below the top-color
scale $\Lambda \sim 1$ TeV the auxiliary fields acquire kinetic terms,
becoming physical degrees of freedom. The properly renormalized
$\phi_1$ and $\phi_2$ doublets take the form
\beq
\phi_1 = \left ( \begin{array}{cc}
\fq + \frac{1}{\sqrt{2}}(h_t + i \pcz) \\ \pcm
\end{array} \right ), \ \
\phi_2 = \left ( \begin{array}{cc}
\tilde{H}^+\\ \frac{1}{\sqrt{2}}(\tilde{H}^0 + i \tilde{A}^0)
\end{array} \right ), \label{phi12}
\eeq
where $\pcc$ and $\pcz$ are the top pions, $\tilde{H}^{\pm,0}$ and
$\tilde{A}^0$ are the b pions, $h_t$ is the top Higgs boson,
and $\fq \approx 50 $ GeV is the top pion decay constant.

From Eq. (\ref{eff2}), the couplings of top pions
to t and b quark can be written as \cite{hill95}
\beq
\frac{m_t^*}{ \fq } \left[ i\; \bar{t} t \tilde{\pi}^0 +
  i\; \overline{t}_R b_L \tilde{\pi}^+
+ i\; \frac{m_b^*}{m_t^*}  \overline{t}_L b_R \tilde{\pi}^+ + H.c.
\right], \eeq
where $m_t^* = (1-\epsilon) m_t$ and $m_b^* \approx 1$ GeV denote the masses
of top and bottom quarks generated by top-color interactions.

For the mass of top pions, the current $1-\sigma$ lower mass bound from the
Tevatron data is $\mpcc \geq 150 $ GeV \cite{lane96}, while the theoretical
expectation is $\mpcc \approx (150 - 300 GeV)$\cite{hill95}. For the mass
of b pions, the current theoretical estimation is $m_{\tilde{H}^0} \approx
m_{\tilde{A}^0} \approx (100 - 350)  GeV$ and $\mpdd = m_{\tilde{H}^0}^2
+ 2 m_t^2$ \cite{kominis95}. For the technipions $\paa$ and
$\pbb$, the theoretical estimations are $\mpaa
\geq 50 GeV$ and $\mpbb \approx 200GeV$\cite{eichten86,epj981}.
The effective Yukawa couplings of ordinary technipions $\paa$ and $\pbb$ to
fermion pairs, as well as  the gauge couplings of unit-charged scalars to gauge
bosons $\gamma, Z^0$ and $gluon$ are basically model-independent,  can be
found in Refs.\cite{eichten86,epj981,ellis81}.

At low energy, potentially large flavor-changing neutral currents (FCNC) arise
when the quark fields are rotated from their weak eigenbasis to their mass
eigenbasis, realized by the matrices $U_{L,R}$ for the up-type quarks,
and by $D_{L,R}$ for the down-type quarks. When we make the replacements,
for example,
\beq
b_L \to  D_L^{bd} d_L  +   D_L^{bs} s_L + D_L^{bb} b_L, \\
b_R \to D_R^{bd} d_R + D_R^{bs} s_R + D_R^{bb} b_R,
\eeq
the FCNC interactions will be induced. In TC2 model, the corresponding
flavor changing effective Yukawa  couplings  are
\beq
\frac{m_t^*}{\fq} \left[
i\; \tilde{\pi}^+ ( D_L^{bs}\bar{t}_R  s_L +  D_L^{bd}\bar{t}_R d_L) +
i\; \tilde{H}^+ ( D_R^{bs} \bar{t}_L s_R +   D_R^{bd}\bar{t}_L d_R)
+ H.c. \right ].
\eeq

For the mixing matrices in the TC2 model, authors usually use the
``square-root ansatz":  to take the
square root of the standard model CKM matrix ($V_{CKM}=U_L^+ D_L$)
as an indication of the size of realistic mixings. It should be denoted that
the square root ansatz must be modified because of the strong constraint from
the data of $B^0 - \overline{B^0}$ mixing \cite{kominis95,epj983,burdman01}.
In the TC2 model, the neutral scalars $\tilde{H}^0$ and $\tilde{A}^0$ can induce
a contribution to the $B_q^0-\overline{B_q^0}$ ($q=d, s$) mass difference
\cite{buchalla96b,kominis95}
\beq
\frac{\Delta M_{B_q}}{M_{B_q}}
= \frac{7}{12}\frac{m_t^2}{\fq^2 m^2_{\tilde{H}^0}}
\delta_{bq}B_{B_q} F_{B_q}^2 ,
\label{deltabd}
\eeq
where $M_{B_q}$ is the mass of $B_q$ meson, $F_{B_q}$ is the $B_q$-meson
decay constant, $B_{B_q}$ is the renormalization group invariant parameter,
and $\delta_{bq} \approx |D_L^{bq}D_R^{bq}|$. For the $B_d$ meson, using the
experimental measurement of $\Delta M_{B_d}=(3.22 \pm 0.05)\times
10^{-10} MeV$\cite{pdg2002} and setting $\fq =45GeV$,
$\sqrt{B_{B_d}}F_{B_d}=200 MeV$,  one has the bound $\delta_{bd} \leq 0.82
\times 10^{-7}$ for $m_{\tilde{H}^0} \leq 600 GeV $.
This is an important and strong bound on the product of mixing elements
$D_{L,R}^{bd}$. As pointed in \cite{buchalla96b},
if one naively uses the square-root ansatz for {\em both}  $D_L$ and $D_R$,
this bound  is violated by about 2 orders of magnitude.
By taking into account above experimental constraint, we naturally set that
$D_R^{ij}=0$ for $i\neq j$. Under this assumption, only the charged
technipions $\paa, \pbb$ and the charged top pions $\pcc$ contribute to the
decays studied here through penguin diagrams.

\subsection{Constraint on TC2 model from $B \to X_s \gamma$ decay}

The constraint on both $D_L$ and $D_R$ from the experimental data of $\bsga$ decay is
much weaker than that from the $B^0-\bar{B}^0$ mixings\cite{buchalla96b}.
On the other hand, one can draw strong constraint on the mass of top-pion $\mpcc$
from the well measured $B \to X_s \gamma $ decay by setting $D_L^{bd}=V_{td}/2$,
$D_L^{bs}=V_{ts}/2$, $\fq=45$ GeV and $\epsilon =0.05\pm 0.03$.

In this subsection, we firstly calculate the new physics contributions to the
Wilson coefficients $C_7(\mw)$ and $C_8(\mw)$. And then we draw the constraint
on the mass $\mpcc$ by comparing the theoretical prediction of ${\calb } (B \to X_s \gamma)$
with the measured value as given in Eq.(\ref{eq:bsgaexp}).

\begin{figure}[t]       
\vspace{-300pt}
\begin{minipage}[]{\textwidth}
\centerline{\epsfxsize=\textwidth \epsffile{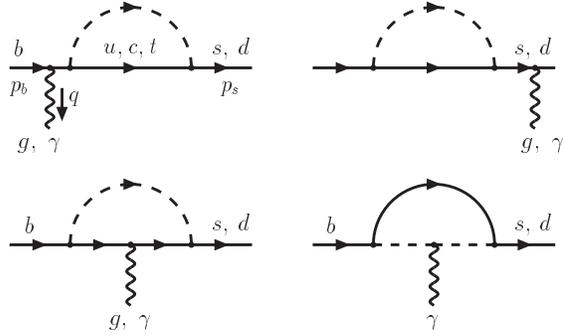}}
\vspace{-330pt}
\caption{The typical photon- and gluon-penguin diagrams with W  and
charged-PGB exchanges (short-dashed lines) in the SM and TC2 models which contribute to
$B \to X_{s,d} \gamma$ decays. The internal quarks are the upper type $u, c$ and $t$
quarks.}
\label{fig:fig3}
\end{minipage}
\end{figure}

The new photonic- and gluonic-penguin diagrams can be obtained from the
corresponding penguin diagrams in the SM by replacing the internal $W^{\pm}$
lines with the unit-charged scalar ($\paa, \pbb$ and $\pcc$ ) lines, as shown in
\fig{fig:fig3}. For details of the analytical calculations, one can see Ref.\cite{epj18}.

By evaluating the new $\gamma$-penguin and $gluon$-penguin diagrams induced by the exchanges
of three kinds of charged pseudoscalars ($\pcc, \paa, \pbb$), we find that
\beq
C_7(\mw)^{TC2} &=& \frac{1}{8\sqrt{2}G_F \fq^2 } H(y_t)
+ \frac{1}{6\sqrt{2}G_F F_\pi }\left [ H(\eta_t) + 8 H(\xi_t)\right ],
\label{eq:c7np}\\
C_8(\mw)^{TC2} &=& \frac{1}{8\sqrt{2}G_F \fq^2 } K(y_t)
+ \frac{1}{6\sqrt{2}G_F F_\pi }\left [ K(\eta_t) + 8 K(\xi_t) + 9 L(\xi_t)\right ],
\label{eq:c8np}
\eeq
where $y_t=\mpcc^2/((1-\epsilon)m_t)^2$, $\eta_t=\mpaa^2/(\epsilon m_t)^2$,
$\xi_t=\mpbb^2/(\epsilon m_t)^2$, while the functions $H(x)$, $K(x)$ and $L(x)$
are
\beq
H(x)&=& \frac{22-53 x + 25 x^2}{36 (1-x)^3} + \frac{3x- 8x^2 + 4 x^3}{6(1-x)^4}\log[x], \label{eq:hx}\\
K(x)&=& \frac{5-19 x + 20 x^2}{12 (1-x)^3} - \frac{x^2-2 x^3}{2(1-x)^4}\log[x], \label{eq:kx}\\
L(x)&=& \frac{4-5x - 5x^2}{12(1-x)^3} + \frac{x-2x^2}{2(1-x)^4}\log[x]. \label{eq:lx}
\eeq
It is easy to show that the charged top-pion $\pcc$ strongly dominate the
new physics contributions to the Wilson coefficients $C_7(\mw)$ and $C_8(\mw)$,
while the technipions play
a  minor rule only, less than $5\%$ of the total NP correction. We therefore
fix the masses of $\paa$ and $\pbb$ in the range of $\mpaa=200\pm 100$ GeV and
$\mpbb=400 \pm 100$ GeV, as listed in \tab{npinput}.
At the leading order, the charged-scalars
do not contribute to the remaining Wilson coefficients $Q_1 -Q_6$.

When the new physics contributions are taken into account, the Wilson coefficients
$C_7(\mw)$ and $C_8(\mw)$ can be defined as the following,
\beq
C_7(\mw)^{Tot}&=& C_7(\mw)^{SM} + C_7(\mw)^{TC2}, \label{eq:c7t}\\
C_8(\mw)^{Tot}&=& C_8(\mw)^{SM} + C_8(\mw)^{TC2}, \label{eq:c8t}
\eeq
where $C_{7,8}^{SM}$ have been given in Eqs.(\ref{eq:c7mw}), (\ref{eq:c8mw}).
Explicit calculations show that the Wilson coefficients $C_{7,8}^{TC2}$ have
the opposite sign with their SM counterparts, and therefore they will
interfere  destructively. The QCD running of $C_{7}^{tot}$ from the energy
scale $\mw$ to $\mu \approx m_b$ is the same as the case of SM.

\begin{figure}[]     
\vspace{-60pt}
\begin{minipage}[]{\textwidth}
\centerline{\epsfxsize=\textwidth \epsffile{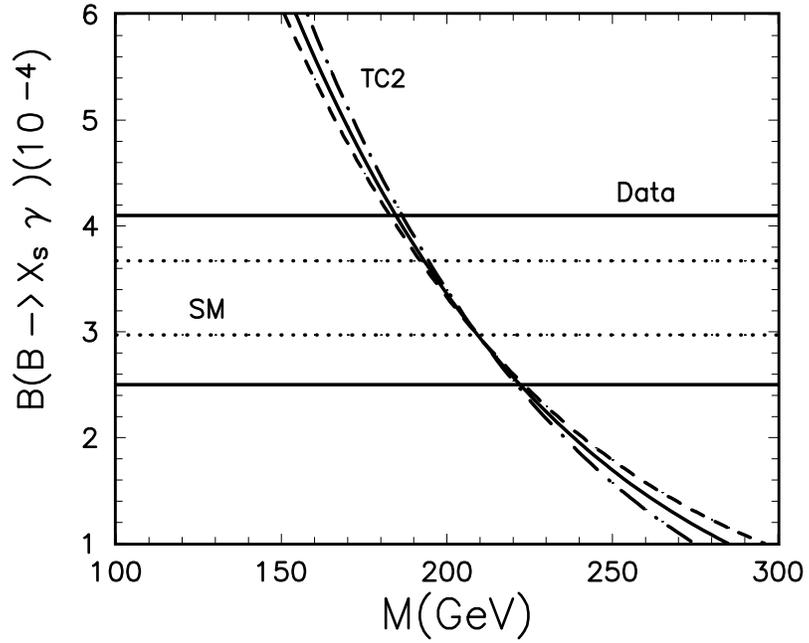}}
\vspace{-30pt}
\caption{The branching ratios ${\calb } (B \to X_s \gamma)$ in the SM and TC2
models as a function of $\mpcc$. The band between two horizontal dashed-lines
(solid lines) shows the SM
prediction (world average of experimental measurements) as listed in
Eqs.(\ref{eq:bsgaexp}), (\ref{eq:bsgath}). The short-dash, solid and dot-dash curves
show the TC2 model predictions of the branching ratios for $\mu=m_b/2, m_b$ and
$2m_b$, respectively.} \label{fig:fig4}
\end{minipage}
\end{figure}

Using the NLO formulas as presented in Ref.\cite{kagan99} for the $B \to X_s
\gamma$ decay, we find the numerical results for the branching ratios
${\calb} (B \to X_s \gamma)$ in both the SM and the TC2 model, as illustrated
in \fig{fig:fig4}, where we use the central values of input parameters as given
in \tab{input} and \tab{npinput}. The three curves correspond to
$\mu=m_b/2$ (short-dashed curve), $\mu=m_b$ (solid curve) and $\mu=2 m_b$ (dot-dashed
curve), respectively. The band between two horizontal dotted-lines  shows the SM
prediction ${\calb}(B \to X_s \gamma)=(3.29 \pm 0.34) \times 10^{-4}$ \cite{kagan99}, while
the band between two horizontal solid lines  shows the data,
$2.5 \times 10^{-4} \leq {\calb}(B \to X_s \gamma) \leq 4.1 \times 10^{-4}$
at the $2\sigma$ level \cite{pdg2002}.

From \fig{fig:fig4} and considering the errors induced by varying $\mpaa$, $\mpbb$ and $\epsilon$
in the ranges as shown in \tab{npinput}, the constraint on the mass of charged
top pion is
\beq
\mpcc= 200 \pm 30 GeV, \label{eq:limit}
\eeq
which is a rather strong constraint on $\mpcc$.

\section{Numerical results in TC2 model} \label{sec:4}

In this section, we present the numerical results for the branching ratios and
CP violating asymmetries of $B_{s,d} \to \gamma \gamma$ decays in the TC2 model.

\subsection{Branching ratios ${\calb}(B_{s,d} \to \gamma \gamma)$ in TC2 model}

Based on the analysis in previous sections, it is straightforward to
present the numerical results. Our choice of input parameters are summarized
in \tab{input} and \tab{npinput}. Using the input parameters as given in \tab{input}
and \tab{npinput} and assuming $\gamma =(60\pm 20)^\circ$,
we find  the numerical results of the branching ratios
\beq
{\calb }(\bar{B}_s \to \gamma \gamma ) &=& \left [2.8 ^{+6.0}_{-1.4}(\Delta \lambda_B)
 ^{+1.3}_{-1.2} (\Delta \mu)  ^{+0.8}_{-0.7}(\Delta f_{B_s}) ^{+1.2}_{-0.8}(\Delta \mpcc)
 \right ] \times 10^{-6},  \label{eq:brbs2}\\
{\calb }(\bar{B}_d \to \gamma \gamma ) &=& \left [8.2 ^{+17.0}_{-4.2} (\Delta \lambda_B)
^{+3.9}_{-3.5} (\Delta \mu) ^{+2.7}_{-2.3} (\Delta f_{B_d}) ^{+3.3}_{-2.3} (\Delta \mpcc)
 \right ] \times 10^{-8}, \label{eq:brbd2}
\eeq
where the major errors correspond to the uncertainties of $\Delta \lambda_B=\pm 0.15$ GeV,
$m_b/2\leq  \mu \leq 2 m_b$, $\Delta f_{B_q} =\pm 0.03$ GeV and $\Delta \mpcc = 30$ GeV,
respectively.

Figures 5(a) and 5(b) show the charged top-pion mass and
$\mu $-dependence of the decay rates ${\calb}(B_{s,d} \to \gamma \gamma )$, respectively.
In these figures, the lower three lines show the SM predictions for
$\mu=m_b/2$(dotted line), $\mu=m_b$(solid line) and $\mu=2 m_b$ (short-dashed
line). Other three curves correspond to the theoretical predictions of TC2 model.
The new physics enhancement on the branching ratios and their scale and mass
dependence can be seen easily from the figure.

From the numerical results as given in Eqs.(\ref{eq:brbs2}), (\ref{eq:brbd2}),
it is easy to see that the largest error of the theoretical prediction comes from
our ignorance of hadronic parameter $\lambda_B$. We show such $\lambda_B$ dependence of
branching ratios in Fig.\ref{fig:fig6} explicitly.
The dotted and short-dashed curves in Fig.\ref{fig:fig6} show the
SM predictions for $\mu=m_b/2$ and $\mu=m_b$, respectively.
The dot-dashed and solid curves show the TC2 model predictions
for $\mu=m_b/2$ and $\mu=m_b$, respectively.
The decay branching ratios decrease quickly, as $\lambda_B$
getting large for both SM and TC2 model.

In order to reduce the errors of theoretical predictions induced by the uncertainties of
input parameters, we define the ratio $R(B_q \to \gamma \gamma)$ with $q=d,s$ as follows
\beq
R(B_{q} \to \gamma \gamma) = \frac{\calb (B_{q} \to \gamma \gamma)^{TC2}}{
\calb (B_{q} \to \gamma \gamma)^{SM}}.\label{eq:ratio}
\eeq
Using the central values of input parameters, one finds numerically that
\beq
R(B_s \to \gamma \gamma) &=& 2.34  \pm 0.10 (\Delta \lambda_B)
 ^{+1.70}_{-1.22} (\Delta \mu)  ^{+0.94}_{-0.68} (\Delta \mpcc),  \label{eq:rbs1}\\
R(B_d \to \gamma \gamma )&=& 2.56 \pm 0.02 (\Delta \lambda_B)
^{+2.10}_{-1.40} (\Delta \mu) ^{+1.01}_{-0.73} (\Delta \mpcc), \label{eq:rbd1}
\eeq
where the errors correspond to $\Delta \lambda_B=\pm 0.15$ GeV,
$m_b/2\leq  \mu \leq 2 m_b$  and $\Delta \mpcc = 30$ GeV, respectively.
The dependence on input parameters $f_B$, $m_B^3$, $G_F$ and $\alpha_{em}$
cancelled in the ratio $R$.

In Figs. 7(a) and 7(b), we show the $\mu$, $\mpcc$ and $\lambda_B$ dependence of the
ratio $R$ explicitly. It is easy to see from \fig{fig:fig7}(b) that the strong
$\lambda_B$-dependence of the individual branching ratios is now greatly reduced in the ratio
$R$, but the strong $\mu$-dependence still remains large.
Obviously, the new physics enhancements to both branching ratios can be as large as a factor
of two to six within the reasonable parameter space.

\subsection{Direct CP violation of $B_{s,d} \to \gamma \gamma$ in TC2 model}

Now we calculate the new physics correction on the CP violating asymmetries of
$B_{s,d} \to \gamma \gamma$ decays. By using the input parameters as given
in Tables I  and III, we find the numerical results as follows:
\beq
r^-_{CP} (B_s \to \gamma \gamma)^{TC2} &=& \left [ -0.25 ^{+0.10}_{-0.06}(\Delta \mu)
\pm 0.10 (\Delta \lambda_B)  \pm 0.04 (\Delta \mpcc ) ^{+0.07}_{-0.03}(\Delta \gamma)
 \right ] \times 10^{-2}, \label{eq:rcpbs2}\\
r^-_{CP} (B_d \to \gamma \gamma)^{TC2 }&=& \left [ + 6.5  ^{+1.7}_{-3.9}(\Delta \mu)
^{+2.8}_{-2.0}(\Delta \lambda_B) \pm 1.1 (\Delta \mpcc)  ^{-0.1}_{-0.9}  (\Delta \gamma )\right ]
 \times 10^{-2}, \label{eq:rcpbd2}
\eeq
where the major errors are induced by the uncertainties of the corresponding
input parameters $\Delta \lambda_B=\pm 0.15$ GeV, $m_b/2\leq  \mu \leq 2 m_b$,
$\Delta \mpcc = 30$ GeV and $\Delta \gamma =\pm 20^\circ $, respectively.

For the $B_s \to \gamma \gamma$ decay, its direct CP violation is still very small
after inclusion of new physics corrections. For the $B_d \to \gamma \gamma$ decay,
however, its CP violating asymmetry is around $7\%$ in TC2 model and depends on
the hadronic parameter $\lambda_B$, the scale $\mu$, the CKM angle $\gamma$
and the mass $\mpcc$, as illustrated by \figs{fig:fig8} and 9.

In \fig{fig:fig8} we draw the plots of the CP violating asymmetry
$r^-_{CP} (B_d \to \gamma \gamma)$
versus the parameters $\mu$, $\lambda_B$ and $\gamma$. The lower  and upper  three curves
in \fig{fig:fig8} show the theoretical predictions of the SM and TC2 model, respectively.
In \fig{fig:fig8}(b), $\gamma=60^\circ$ is assumed.
It is easy to see from \fig{fig:fig8} that the pattern of the CP violating asymmetry in TC2 model
is very different from that in the SM.
The sign of $r^-_{CP} (B_d \to \gamma \gamma)$ in TC2 model is opposite to that  in the SM,
while its size does not change a lot. Such difference can be detected when the statistics of
the current and future B experiments becomes large enough.

\section{Discussions and summary}

In this paper, we calculate the new physics contributions to the branching ratios
and CP-violating asymmetries of double radiative decays $B_{s,d} \to \gamma \gamma$
in the TC2 model by employing the QCD factorization approach.

In Sec. \ref{sec:sm}, based on currently available studies, we present the effective
Hamiltonian for the inclusive $\bsga$ and $\bsgg$ decays. For the evaluation of
hadronic matrix elements for the exclusive $B_{s,d} \to \gamma \gamma$ decays, we use
Bosch and Buchalla approach to separate and calculate the leading and subleading
power contributions to the exclusive decays under study from 1PR and 1PI Feynman diagrams.
We  reproduce the SM predictions for the branching ratios $\calb (B_{s,d} \to \gamma \gamma)$
and direct CP asymmetries $\rcppm$ as given in Ref.\cite{bosch02a}.

For the new physics part, we firstly give a brief review about the basic structure of TC2 model,
and evaluate analytically the strong and electroweak charged-scalar
penguin diagrams in the quark level processes $b \to s/d \gamma $ and $ b \to s g$. We
extract out the new physics contributions to the corresponding
Wilson coefficients $C_7(\mw)$ and $C_8(\mw)$. Then we
combine these new functions with their SM counterparts and run these Wilson coefficients
from the scale $\mu=\mw $ down to the lower energy scale $\mu =O(m_b)$ by using
the QCD renormalization equations.
From the data of $B_d^0-\bar{B}_d^0$ mixing, we find the strong constraint on
the ``square-root ansatz". We also extract the strong constraint on the mass $\mpcc$ by comparing
the theoretical predictions for the branching ratio $\calb (B\to X_s \gamma)$ at the NLO level
with the experimental measurements.

\begin{figure}[t] 
\vspace{-100pt}
\begin{minipage}[t]{\textwidth}
\centerline{\epsfxsize=\textwidth \epsffile{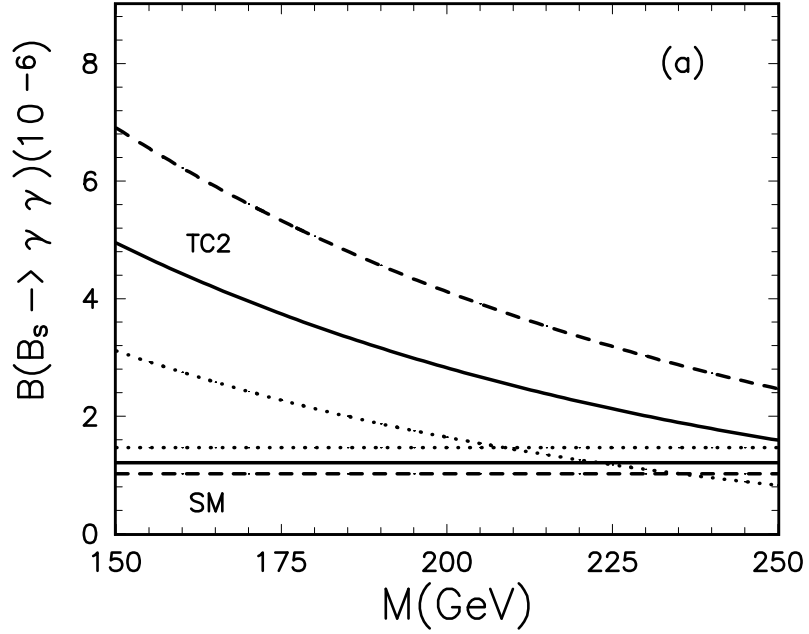}}
\vspace{-60pt}
\centerline{\epsfxsize=\textwidth \epsffile{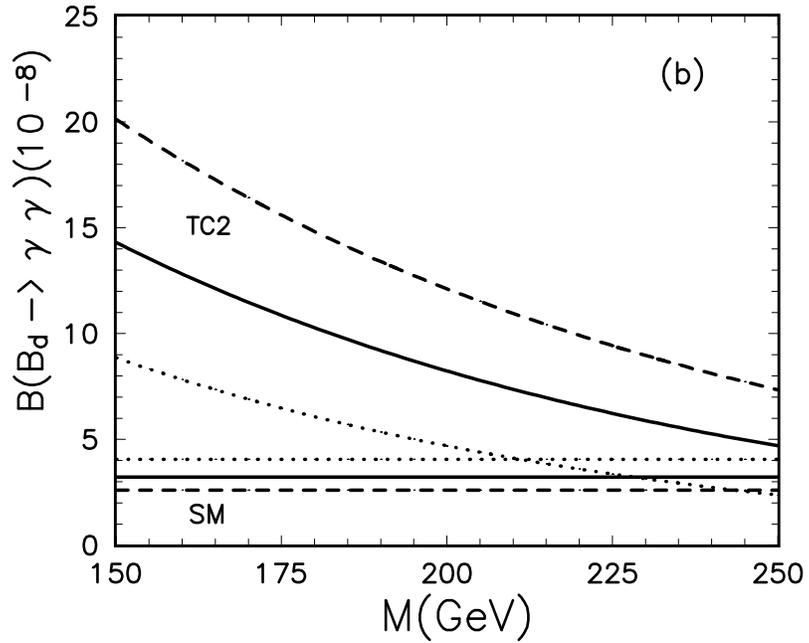}}
\vspace{-20pt}
\caption{Plots of branching ratios ${\calb}(B_{s} \to \gamma \gamma)$ (a) and
${\calb}(B_{d} \to \gamma \gamma)$ (b) vs $\mpcc$,
setting $\lambda_B=0.35$ and CKM angle $\gamma=60^\circ$.
The lower three lines in each diagram show the SM predictions for
$\mu=m_b/2$ (dotted line), $\mu=m_b$ (solid line) and $\mu=2 m_b$ (short-dashed
line). Upper three curves correspond to the theoretical predictions of TC2 model.}
\label{fig:fig5}
\end{minipage}
\end{figure}

\begin{figure}[t] 
\vspace{-100pt}
\begin{minipage}[t]{\textwidth}
\centerline{\epsfxsize=\textwidth \epsffile{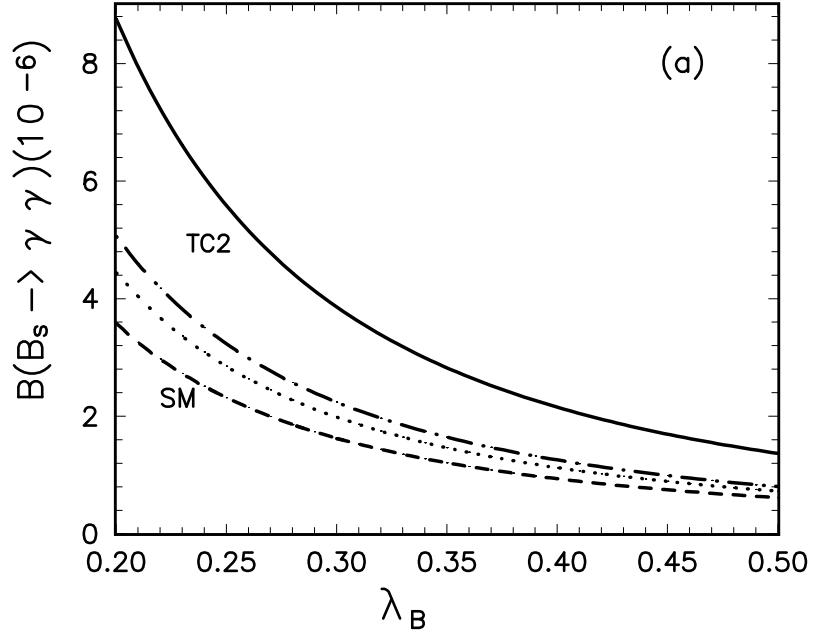}}
\vspace{-60pt}
\centerline{\epsfxsize=\textwidth \epsffile{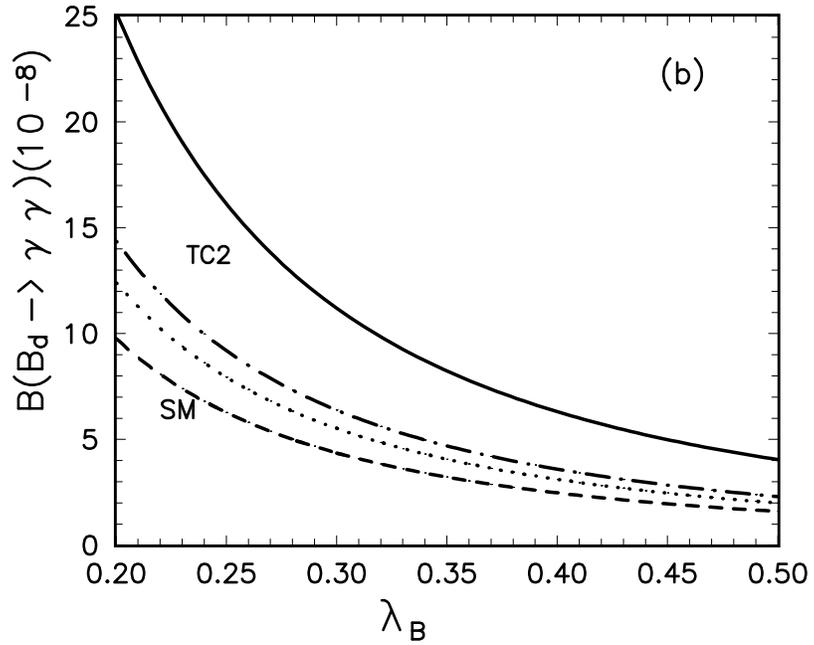}}
\vspace{-20pt}
\caption{Plots of branching ratios ${\calb}(B_{s} \to \gamma \gamma)$ (a) and
${\calb}(B_{d} \to \gamma \gamma)$ (b) vs $\lambda_B$,
setting $\mpcc=200$ GeV and CKM angle $\gamma=60^\circ$.
The dotted and short-dashed curves show the SM predictions
for $\mu=m_b/2$ and $\mu=m_b$, respectively.
The dot-dashed and solid curves show the TC2 model predictions
for $\mu=m_b/2$ and $\mu=m_b$, respectively. }
\label{fig:fig6}
\end{minipage}
\end{figure}

\begin{figure}[t] 
\vspace{-100pt}
\begin{minipage}[t]{\textwidth}
\centerline{\epsfxsize=0.95\textwidth \epsffile{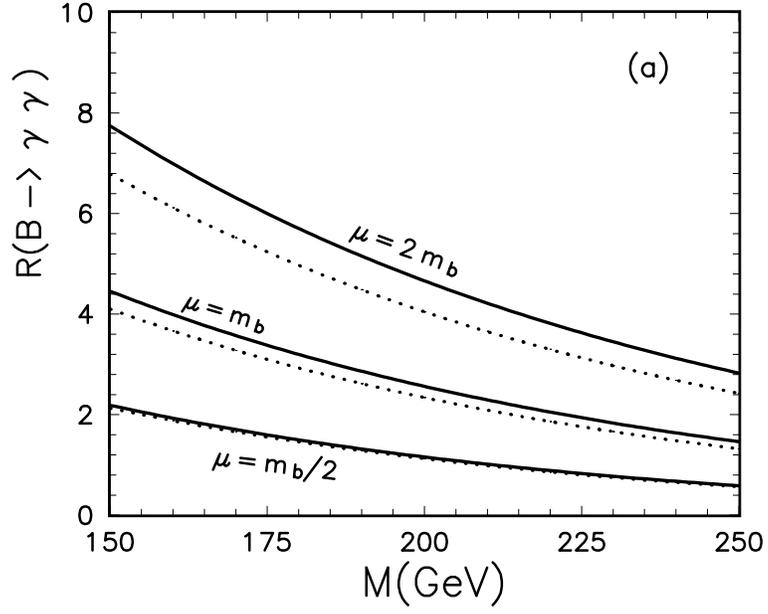}}
\vspace{-60pt}
\centerline{\epsfxsize=\textwidth \epsffile{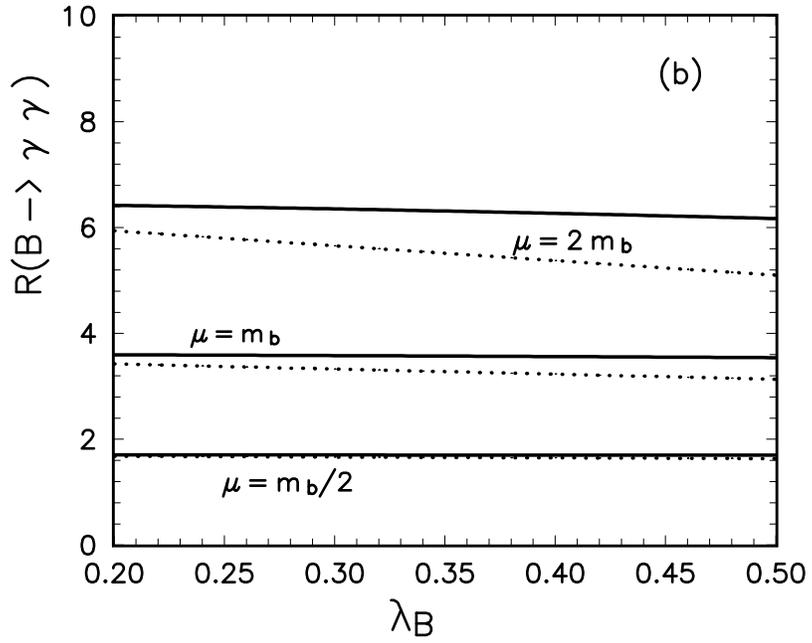}}
\vspace{-20pt}
\caption{The ratio of branching ratios $R(B_{s,d} \to \gamma \gamma)$ in
the TC2 model. The three dotted and  three solid curves
show the ratios for $B_s \to \gamma \gamma$  and $B_d \to \gamma \gamma$ decays,
respectively. In (b), we set $\mpcc=170$ GeV.}
\label{fig:fig7}
\end{minipage}
\end{figure}

\begin{figure}[t]
\vspace{-100pt}
\begin{minipage}[t]{\textwidth}
\centerline{\epsfxsize=\textwidth \epsffile{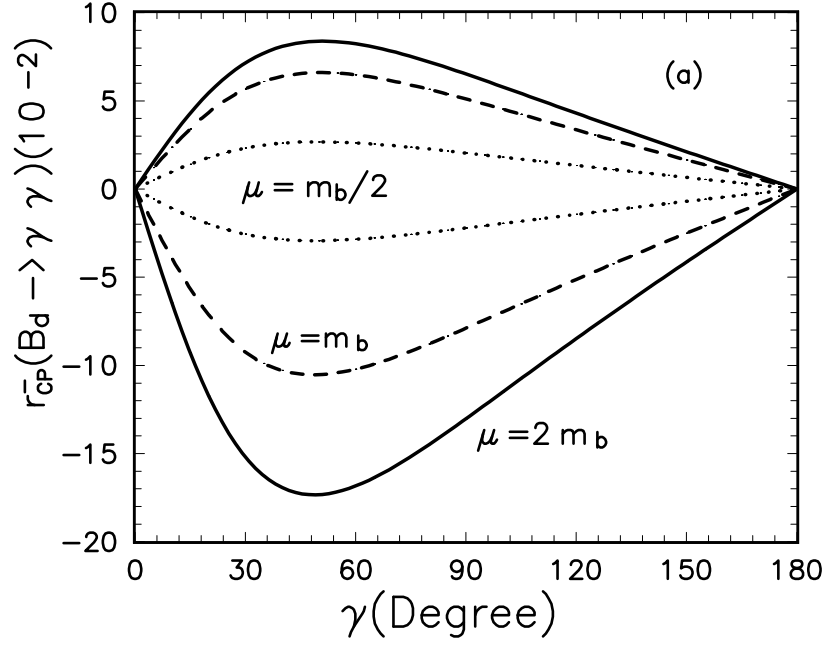}}
\vspace{-60pt}
\centerline{\epsfxsize=\textwidth \epsffile{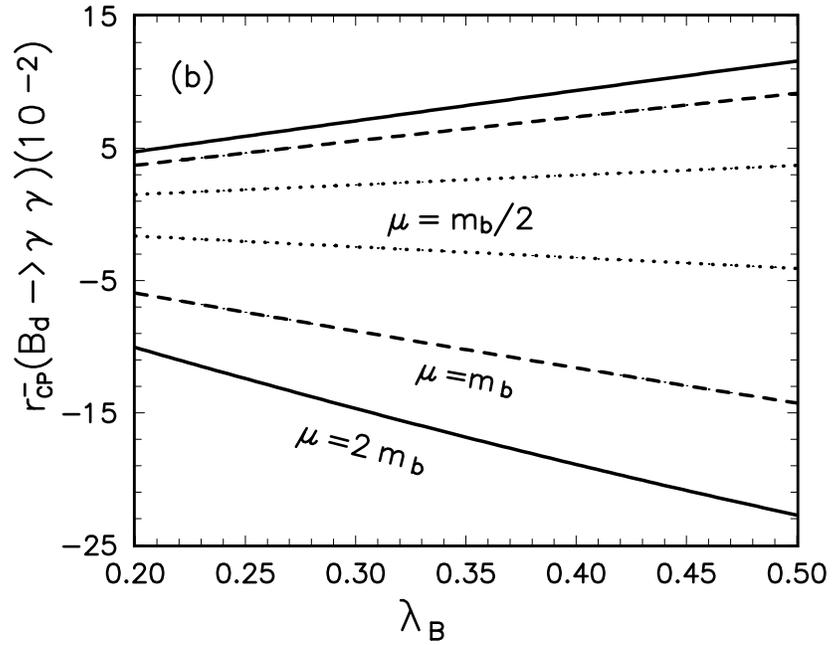}}
\vspace{-20pt}
\caption{The CP violating asymmetry of $(B_d \to \gamma \gamma)$ decay  in
the SM and TC2 model. The lower (upper) dotted, short-dashed and  solid curves
show the SM (TC2) predictions for $\mu=m_b/2, m_b$ and $2m_b$, respectively.}
\label{fig:fig8}
\end{minipage}
\end{figure}

In Sec. \ref{sec:4}, we present the numerical results
for $\calb (B_{s,d} \to \gamma \gamma)$ and $\rcpm(B_{s,d} \to \gamma \gamma)$
after the inclusion of new physics contributions in the TC2 model.
\begin{enumerate}
\item
For both $B_s\to \gamma \gamma$ and $B_d \to \gamma \gamma$ decays, the new
physics contribution can provide a factor of two to six enhancement to their branching ratios.
The $\mpcc$, $\mu$ and $\lambda_B$ dependences are also shown in
\fig{fig:fig5}. With an optimistic choice of the input parameters, the branching ratio
$\calb (B_s \to \gamma \gamma)$ and $\calb (B_d \to \gamma \gamma)$ in the TC2 model  can
reach $10^{-5}$ and $10^{-7}$ respectively, only one order away from the experimental limit
as given in Eqs.(\ref{eq:cleo95}), (\ref{eq:babar01}).
With more integrated luminosity accumulated by BaBar and Belle Collaborations, the upper
bound on $\calb (B_d \to \gamma \gamma)$ will be further improved, and may
reach the interesting region of TC2 prediction.

\item
For the $B_s \to \gamma \gamma$ decay, its direct CP violation is very small
in both the SM and TC2 model.

\item
For the $B_d \to \gamma \gamma$ decay, however, its CP violating asymmetry
is around ten percent level in both the SM and Tc2 model. But the pattern
of CP violating asymmetry in TC2 model is very different from that in the SM,
as illustrated in \fig{fig:fig8}.

\end{enumerate}

As discussed in Ref.\cite{slac8970}, the high luminosity option SuperBaBar suggests a total
integrated luminosity of 10 ab$^{-1}$.  For the branching ratio as given in Eq.(\ref{eq:brbd2}),
the number of observed $B_d \to \gamma \gamma$ events is then expected to be in the range of
$50- 150$ in the TC2 model, and therefore measurable in the future.

\vspace{1cm}

\section*{ACKNOWLEDGMENT}

Z.J.~Xiao acknowledges the support by the National Science Foundation of
China under Grants No.~10075013 and 10275035, and by the Research Foundation of
Nanjing Normal University under Grant No.~214080A916.
C.D.L\"u acknowledges the support  by National Science Foundation of
China under Grants No.~90103013 and 10135060.
W.J.Huo acknowledges  supports from the Chinese Postdoctoral  Science Foundation
and CAS K.C. Wong Postdoctoral Research Award  Fund.

\newpage
\begin{table}
\begin{center}
\caption{Values of the input parameters used in the numerical calculations. All masses
are in units of GeV.}
\label{input}
\begin{tabular}{c c c c c c }
$ A $    & $\lambda $ & $R_b  $         & $G_F$                           & $\alpha_{em}$
    & $\alpha_s(M_Z)$ \\  \hline
$ 0.847$ & $ 0.2205 $ & $0.38\pm 0.08 $ & $1.1664\times 10^{-5}GeV^{-2} $ & $ 1/137.036$
    & $ 0.118 $  \\ \hline \hline
$ m_W  $ & $m_t  $ & $m_b^{pole} $        & $m_c^{pole} $ &$m_{B_d} $ &$ m_{B_s}$ \\  \hline
$ 80.42$ & $ 175 $ & $4.80\pm 0.15  $ & $1.4\pm 0.12$ & $5.279 $ & $5.369 $  \\
\hline\hline
$ f_{B_d}    $ & $f_{B_s} $ & $\lambda_{B_s}=\lambda_{B_d} $ & $\Lambda^{(5)}_{\overline{MS}}$
&$\tau(B_d)$ &$ \tau(B_s)$ \\  \hline
$0.20\pm 0.03$ & $0.23\pm 0.03 $ & $0.35 \pm 0.15 $ & $ 0.225$ & $ 1.542 ps$ & $1.461 ps $
\end{tabular} \end{center}
\end{table}

\begin{table}
\begin{center}
\caption{The LO Wilson coefficients  $C_i(\mu)$ in the SM obtained by using  the central
values of input parameters as listed in Table I. } \label{cieff}
\begin{tabular}{c c c c c c c c c}
 $\mu $ &  $C_1 $  & $C_2 $  & $C_3$ & $C_4$ & $C_5 $ & $C_6 $ & $ C_7$& $C_8$ \\  \hline
$m_b /2$  & $-0.3500$ & $1.1630 $ & $0.0164 $ & $ -0.0351 $ & $0.0096 $ & $-0.0467 $& $ -0.3545$ &$-0.1649$  \\ \hline
$m_b   $  & $-0.2454$ & $1.1057 $ & $0.0109 $ & $ -0.0254 $ & $0.0073 $ & $-0.0309 $& $ -0.3141$ &$-0.1490$
 \\ \hline
$2 m_b $  & $-0.1654$ & $1.0664  $ & $0.0070 $ & $ -0.0175$ & $0.0052
$& $-0.0200 $& $ -0.2801$ &$ -0.1353$
\end{tabular} \end{center}
\end{table}

\begin{table}
\begin{center}
\caption{Values of the input parameters of TC2 model. All masses are in units of GeV.}
\label{npinput}
\begin{tabular}{c c c c c c }
$ \mpaa $     & $\mpbb  $      & $\mpcc $     & $F_\pi $ &$\fq $ &$\epsilon $     \\  \hline
$ 200\pm 100$ & $ 400\pm 100 $ & $200\pm 30 $ & $120$    & $45 $   & $0.05 \pm 0.03$
\end{tabular} \end{center}
\end{table}


\newpage

\newpage

\newpage

\newpage

\newpage

\newpage

\newpage

\begin{figure}[t]
\vspace{-100pt}
\begin{minipage}[t]{\textwidth}
\centerline{\epsfxsize=\textwidth \epsffile{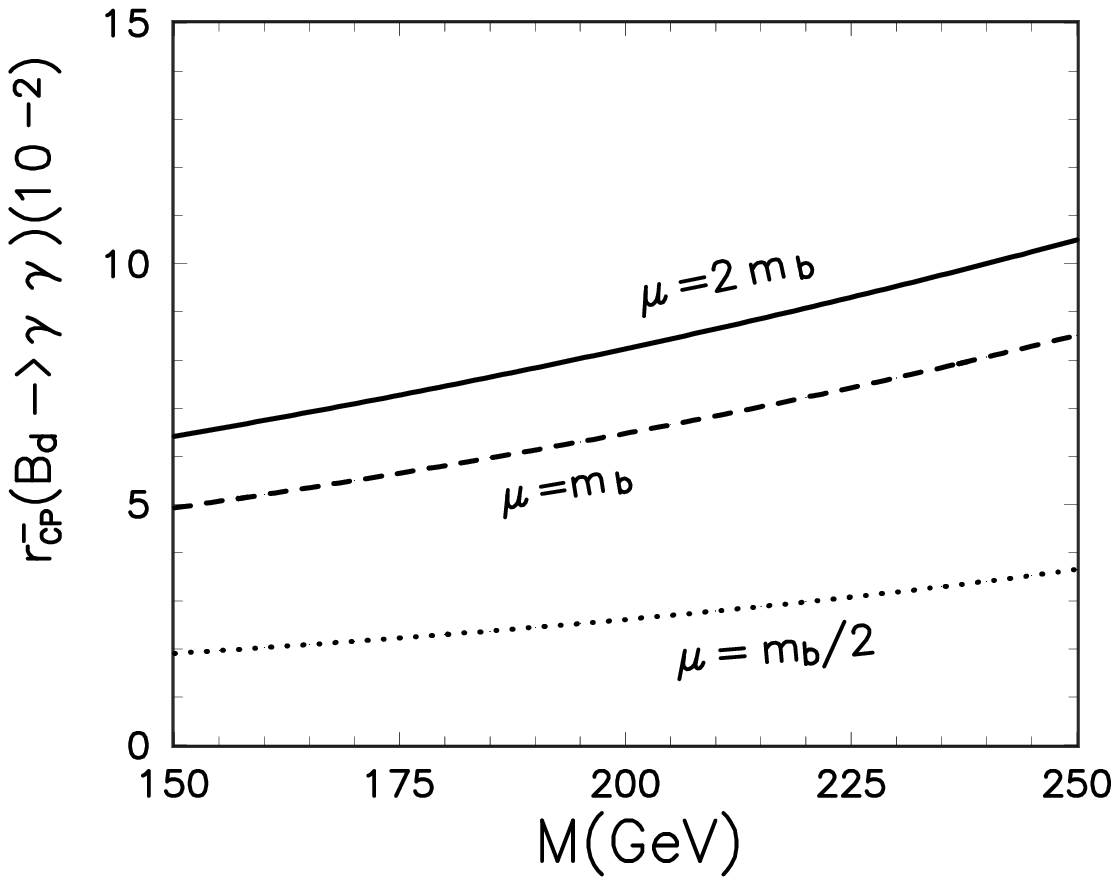}}
\vspace{-20pt}
\caption{The CP violating asymmetry of $(B_d \to \gamma \gamma)$ decay vs
mass $\mpcc$ and energy scale $\mu$ in TC2 model. The dotted, short-dashed and  solid curves
show the TC2 predictions for $\mu=m_b/2, m_b$ and $2m_b$, respectively.}
\label{fig:fig9}
\end{minipage}
\end{figure}

\end{document}